\documentclass[12pt]{iopart}

\usepackage{graphicx}
\begin{document}

\title[Y. G. Ma, $\phi$ meson production and partonic collectivity at RHIC]{$\phi$ meson production and partonic collectivity at RHIC}

\author{Y. G. Ma \footnote[1]{E-mail: ygma@sinap.ac.cn}}
\address{ Shanghai Institute of Applied Physics, Chinese Academy of Sciences, \\
        P. O. Box 800-204, Shanghai 201800, China}

\begin{abstract}
New results on $\phi$-meson production and elliptic flow $v_{2}$
measurements from RHIC 2004 run (Run-IV) have been reviewed. In
addition, the di-hadron correlation function between the trigged
$\phi$ and $\Omega$ and the associated soft particles was
simulated. Knowledge about these results are discussed.
\end{abstract}


\section{INTRODUCTION}

The primary goal of the Relativistic Heavy-Ion Collider (RHIC) is
to produce and study a state of high density deconfined nuclear
matter called the Quark-Gluon Plasma (QGP). Measurements of the
production of $\phi$-meson created at RHIC Au+Au collisions can
provide insight on the dense matter created in the collisions.
Because $\phi$-meson is expected to have a small cross-section for
interactions with other non-strange particles, and its life time
is relatively long ($\sim$41 fm/$c$), it may keep information of
the early stage of the system's evolution. The observation of
meson and baryon grouping in the nuclear modification factor
($R_{CP}$) and elliptic flow ($v_{2}$) measurements at
intermediate transverse momentum ($p_{T}$) \cite{STAR-K0L} has
been interpreted as a manifestation of bulk partonic matter
hadronization through multi-parton dynamic such as
recombination/coalescence of partons \cite{REC1,REC2,REC3}.
$\phi$-meson provides unique sensitivity to test these theoretical
scenarios, since $\phi$-meson has a mass heavier than proton and
close to the Lambda while it is a meson. The hypothesis of kaon
coalescence as a dominant source for $\phi$-meson production has
been ruled out from the STAR 2002 measurements \cite{STAR-phi}. If
$\phi$ flows, it may be a strong signal of the partonic
collectivity formed at RHIC. Based upon these reasons, the studies
on the property of $\phi$-meson at RHIC energies is of great
interest.

\section{TRANSVERSE MOMENTUM DISTRIBUTION}

\begin{figure}[htb]
\centerline{\includegraphics[scale=0.4]{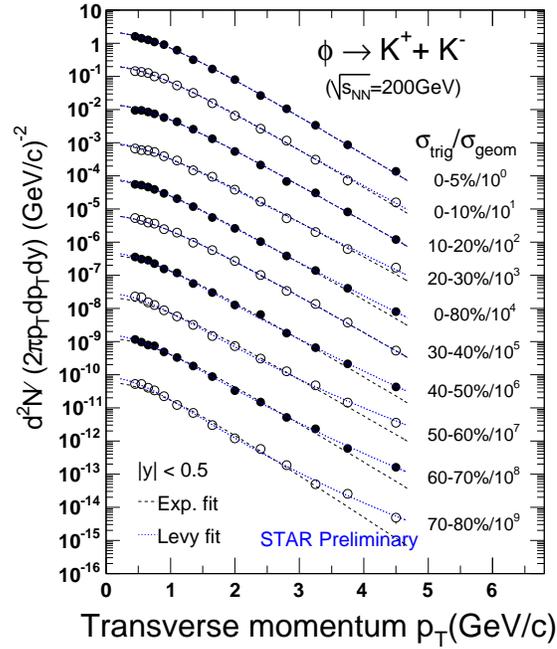}}
\caption{The $\phi$-meson transverse momentum distributions from
Au+Au collisions at $\sqrt{s_{NN}} = 200$ GeV. For clarity,
distributions for different centralities are scaled by factors
which are shown in the figure. Dashed lines represent the
exponential fits to the distributions and dotted lines are the
results of Levy function fits. Error bars represent statistic
errors only.\label{fig1}}
\end{figure}
\begin{figure}[htb]
\centerline{\includegraphics[scale=0.4]{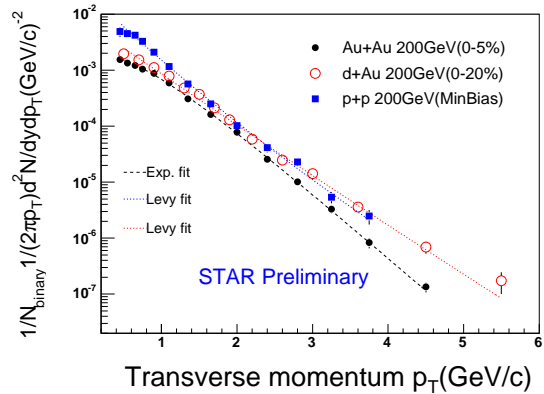}}
\caption{ Comparison of the binary scaled $P_T$ spectra among
different collision systems (Au+Au, d+Au and pp at $\sqrt{s_{NN}}
= 200$ GeV) in their most central centrality bins.\label{fig2}}
\end{figure}

Insight into particle production can be gained by examining the
shape of the $\phi$-meson $p_{T}$ distributions as a function of
centrality \cite{Rudy-Hwa}. Figure~\ref{fig1} shows the
$\phi$-meson transverse momentum distributions as a function of
centralities for Au + Au collisions (RunIV) at $\sqrt{s_{NN}} =
200$ GeV \cite{CAI,Sarah}. It is observed that the shape of the
spectra evolves as a function of centrality. The central data can
be fitted equally well by both exponential function and Levy
functions, while the more peripheral spectra can only be fitted
better by the Levy function. Comparing with the peripheral data,
the yield at high-$p_{T}$ region for central data is suppressed.
Figure~\ref{fig2} \cite{CAI} shows the comparison among different
reaction systems (Au+Au at 200 GeV/c, d+Au at 200 GeV/c and p+p at
200 GeV/c) in their most central bins. The fact that the
$N_{binary}$ normalized spectra of Au+Au collisions at 200 GeV is
an exponential shape and those of d+Au and p+p at 200 GeV/c are
Levy shapes may indicate different reaction mechanism. The
exponential shape at central 200GeV/c Au+Au data may be hints of
thermalization of the system too.

\begin{figure}[htb]
\vspace*{-0.1in}
\centerline{\includegraphics[scale=0.60]{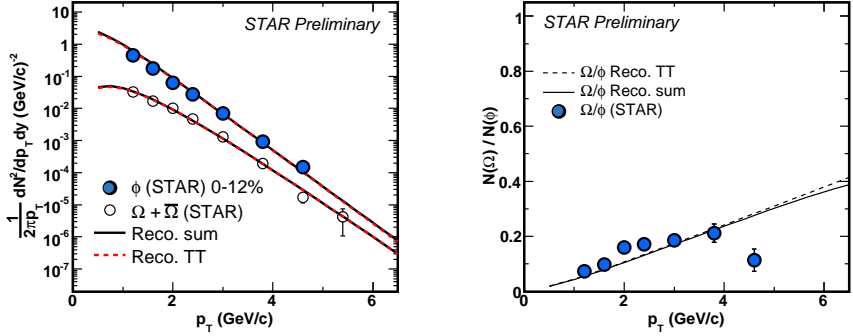}}
\caption{ The left panel shows the $\phi$ and $\Omega$ $p_{T}$
spectra for most central collisions compared to the  recombination
model predictions \cite{Rudy-Hwa}. The sum of all contributions to
the model is shown as the solid lines and the thermal-thermal (TT)
recombination is indicated by the dashed lines. The right panel
shows the $\Omega/\phi$ ratio as a function of $p_{T}$ compared to
the recombination model expectation from the sum of all
contributions (solid line) and the thermal-thermal contribution
alone (dashed line).}\label{fig3}
\end{figure}

Recently, it was predicted \cite{Rudy-Hwa} that the dominant
production mechanism for $\phi$-meson, (and similarly for the
multi-strange baryon, $\Omega$), in the range 2$ < p_{T}<$ 8 GeV/c
is the thermal recombination of strange quarks while the
contribution from fragmentation processes in this $p_{T}$ range is
suppressed by a few orders of the magnitude in comparison. Left
panel of Figure~\ref{fig3} \cite{Sarah} shows the central
$\phi$-meson and $\Omega$ spectra compared with this model
calculation (solid and dashed curves). The model appears to agree
well with the data over the measured $p_{T}$ range. The right
panel of Figure~\ref{fig3} shows the ratio of central $\Omega$
yield to $\phi$-meson yield. In this case, the model predicts a
monotonic increase with $p_{T}$ up to $\sim$ 8 GeV/c. It is
clearly seen that the model describes data well with $p_{T}$ up to
$\sim$ 4 GeV/c while the deviation at higher $p_{T}$ range may
imply that the contribution to particle production from
fragmentation processes is larger than that expected by the model.

\begin{figure}[htb]
\centerline{\includegraphics[scale=0.3]{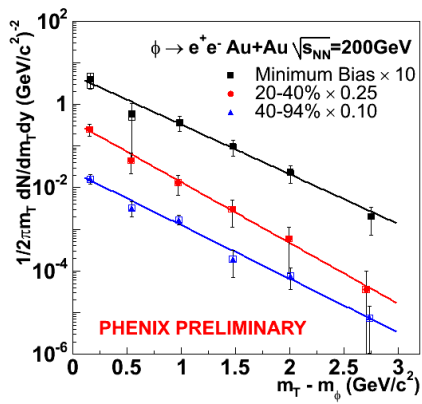}}
\caption{Invariant $m_{T}$ spectra. Statistical and systematic
errors are shown by vertical bars and open rectangles,
respectively \cite{PHENIX-QM05}. \label{fig4}}
\end{figure}

PHENIX presented result of $\phi$-meson production via its
$e^{+}e^{-}$ decay channels (Figure~\ref{fig4}) at Quark Matter
2005 Conference \cite{PHENIX-QM05}. Although the statistic is
still limited, no significant discrepancy between $\phi
\rightarrow e^+e^-$ and $\phi \rightarrow K^+K^-$ has been
observed. In recent CERES results, $\phi \rightarrow e^+e^-$ and
$\phi \rightarrow K^+K^-$ also agree within errors \cite{milo}.
The results agree with NA49's \cite{NA49}, but disagree with
NA50's \cite{NA50}. Thus there is still something for $\phi$
production to be learned. In this case, the di-electrons channel
for $\phi$-meson production measurement from STAR will be of great
interest. This measurement will be achieved with the planned
upgrade of the STAR detector with a full-barrel Time of Flight
detector (TOF) which is expected to significantly increase the
particle identification capability \cite{Zhangbu}.

\vspace{-0.2cm}
\section{NUCLEAR MODIFICATION FACTOR ($R_{CP}$)}
\vspace{-0.2cm}

\begin{figure}[htb]
\vspace*{-0.1in}
\centerline{\includegraphics[scale=0.4]{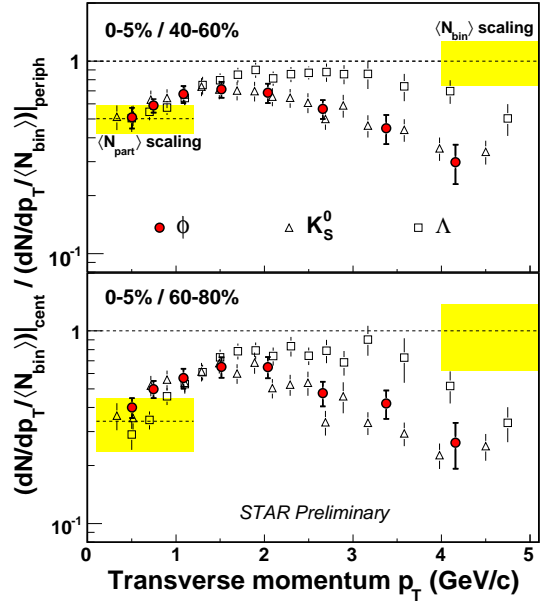}}
\caption{The nuclear modification factor $R_{CP}$ of $\phi$-meson
compared with $R_{CP}$ of $\Lambda$ and $K_{s}^{0}$ at
$\sqrt{s_{NN}} = 200GeV$ \cite{STAR-K0L}. Both statistic and
systematic uncertainty have been included. The shaded bands
represent the uncertainties in the Glauber model calculations for
$\langle$$N_{bin}$$\rangle$ and
$\langle$$N_{part}$$\rangle$.\label{fig5}}
\end{figure}

$R_{CP}$ is calculated as the ratio of the yields from central
collisions to peripheral collisions scaled by the mean number of
binary collisions. Comparisons of the $R_{CP}$ for Au+Au
collisions ($K_{s}^{0}, \phi$ and $\Lambda$) at 200 GeV/c are
shown in Figure~\ref{fig5} \cite{CAI,Sarah}.

From the figure, we can see that at intermediate $p_{T}$, the
$R_{CP}$ of baryon ($\Lambda$) is larger than that of mesons
($K_{s}^{0}$ and $\phi$), which implies that particle production
in this $p_{T}$ region is driven by the particles' types, not by
their masses since $\phi$-meson's mass is close to $\Lambda$ while
it is a meson. The $R_{CP}$ results are consistent with the
partonic recombination model predictions \cite{REC1,REC2,REC3}
that the centrality dependence of the yield at intermediate
$p_{T}$ depends more strongly on the number of constituent quarks
than on the particle mass. There also may be a tendency for values
of $R_{CP}$ for all particles to approach each other at higher
$p_{T}$.

\vspace{-0.2cm}
\section{ELLIPTIC FLOW OF $\phi$ IN LOW AND MIDDLE $p_T$ FOR 200 GeV/c  Au+Au }
\vspace{-0.2cm}

In non-central Au+Au collisions, the overlap region is anisotropic
(nearly elliptic). Large pressure built up in the collision center
results in pressure gradient which is dependent on azimuthal
angle, which generates anisotropy in momentum space, namely
elliptic flow. Once the spatial anisotropy disappears due to the
anisotropic expansion, development of elliptic flow also ceases.
This kind of self-quenching process happens quickly, therefore
elliptic flow is primarily sensitive to the early stage equation
of state (EOS) \cite{elliptic-flow}.
\begin{figure}[htb]
\vspace*{-0.1in}
\centerline{\includegraphics[scale=0.6]{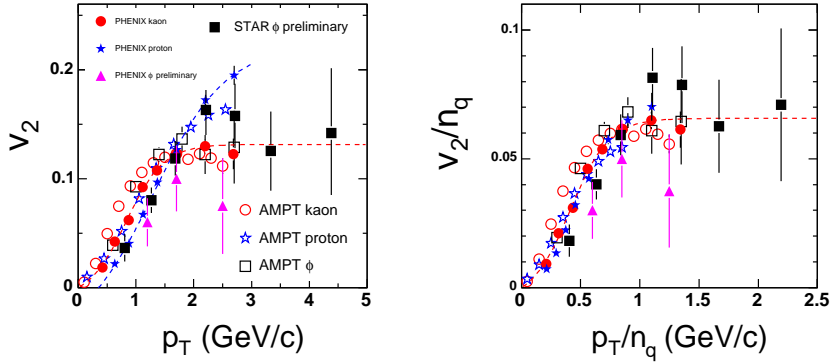}}
\vspace{-0.1cm}
 \caption{Left panel: $p_T$ dependence of $v_2$ for
$\phi$-meson which is compared with the results of $K^{+}$+$K^{-}$
and $p+\overline{p}$. AMPT calculation are taken from
Ref.~\cite{AMPT}.  Right panel: NCQ scaled $v_{2}$ as a function
of NCQ scaled transverse momentum. Lines represent NCQ-scaling
parameterization. \label{fig6}}
\end{figure}

Figure~\ref{fig6} (left panel) shows the preliminary results for
the first measurement of $\phi$-meson elliptic flow $v_{2}$ at
RHIC \cite{CAI,Sarah,PHENIX-phi}.  $v_{2}$ of $K^{+}$+$K^{-}$ and
$p+\overline{p}$ from \cite{PHENIX1} are plotted together for
comparison. The most striking thing in the figure is that the
$\phi$-meson has a non-zero $v_{2}$ in the hydrodynamic region
($p_{T} < 2$ GeV/$c$) and it has a significant $v_{2}$ signal at
$p_{T} > 2$ GeV/$c$ which is comparable to that of the
$K^{+}$+$K^{-}$. Since the formation of $\phi$-meson through kaon
coalescence at RHIC has been ruled out by previous STAR
measurements \cite{STAR-phi}, and the low interaction
cross-section of $\phi$-meson with other non-strange particles
makes the contribution to flow due to hadronic re-scattering
processes very small \cite{AMPT}, therefore it may be possible to
directly measure the flow of strange quark via the flow of
$\phi$-meson. Additionally, since the $\phi$-meson $v_{2}$ values
as a function of $p_{T}$ are similar to those of other particles,
this indicates that the heavier $s$-quarks flow as much as the
lighter $u$ and $d$ quarks. This can happen if there are a
significant number of interactions between the quarks before
hadronization. It is the signal of partonic collectivity of the
system. A prediction \cite{AMPT} from the naive coalescence model
which is implemented in A Multi-Phase Transport (AMPT) model with
the string melting scenario \cite{AMPT2} is also plotted here. The
model describes the data quite well at $p_{T} > 1.5$ GeV/c. This
is another hint of coalescence of strange quark as dominant
production mechanism for $\phi$-meson. Further evidence can be
obtained from the Number-of-Constituent-Quark (NCQ) picture in
Figure~\ref{fig6} (right panel).

\vspace{-0.2cm}
\section{DI-HADRON CORRELATION TRIGGED BY HIGH $p_T$ $\phi$ or $\Omega$}
\vspace{-0.2cm}

The strong suppression of high-$p_{T}$ particle yield
~\cite{hiptsuppress} and the disappearance of one jet in
back-to-back jet correlation~\cite{hard-hard-ex} have been
observed in RHIC Au + Au central collisions at $\sqrt{s_{NN}}$ =
200 GeV, which can be interpreted by jet quenching mechanism
~\cite{HIJING}. On the other hand, the lost energy will be
redistributed in the soft $p_T$ region. These soft associated
particles which carry the lost energy have been reconstructed via
two-particle azimuthal angular correlation of charged particles in
STAR experiment \cite{soft-soft-ex}, and the interesting Mach-like
structure has been experimentally observed in Au + Au collisions
at $\sqrt{s_{NN}}$ = 200 GeV ~\cite{sideward-peak1,sideward-peak3}
as well as theoretically interpreted. For instance, it was
proposed that a Mach shock wave
 happens when jet travels faster than sound in the medium
\cite{Stocker,Casalderrey1,Ruppert}, or it can also be produced
with a Cherenkov radiation mechanism \cite{Koch}, or was
attributed to medium dragging effect in Ref.~\cite{Armesto} etc.
In addition, partonic transport model is able to give the
Mach-like cone phenomenon for central Au+Au collisions at RHIC and
shows the parton cascade is essential for describing experimental
Mach-like structure \cite{MAGL}.

\begin{figure}[htb]
\vspace*{-2.in}
\centerline{\includegraphics[scale=0.55]{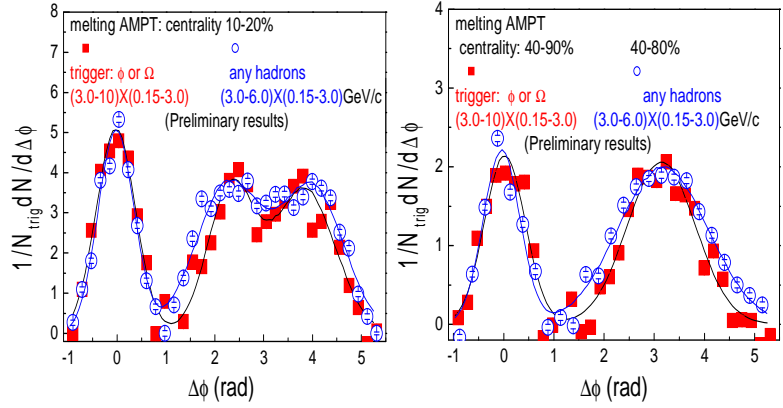}}
\vspace*{-2.2in}\caption{\label{dihadron}Soft scattered associated
hadron ( $0.15 < p_T < 3$ GeV/c ) $\Delta\phi$ correlations
trigged by high $p_T$ particles. Solid squares represent the
trigger particles are either $\phi$ or $\Omega$ with $3 < p_T <
10$ GeV/c, and open circles any hadrons with $3 < p_T < 6$ GeV/c.
Left panel: 10-20 $\%$ centrality; Right panel: 40-90 or 40-80$\%$
centrality. }
\end{figure}

In this section, we shall present two-particle correlation
function, especially for those trigged by multi-strange particles,
such as $\phi$ and $\Omega$ with higher $p_T$ which is beyond the
$p_T$ region we discussed in the section of $v_2$, to further
understand behavior of $\phi$ mesons. Fig.~\ref{dihadron} shows
the di-hadron correlation functions in 10-20$\%$ centrality (left
panel) or 40-80 or 40-90$\%$ centrality (right panel) for the
results before the hadronic rescattering stage takes place in AMPT
model. The mixed-event technique was used to extract the di-hadron
correlation function as people made in RHIC data analysis. The
full squares represent the trigger particles are multi-strange
particles, namely $\phi$ or $\Omega$, while the open circles show
the trigger particles are any hadrons. A clean split of the
away-side peaks is observed in left panel, which is similar to the
Mach-like cone behavior. It is possible to induce such kind of
picture if the jet velocity is faster than the sound velocity when
it passes through the dense matter. Also the deflect may occur
when the associated particles try to traverse the dense QCD
medium.  From the comparison of different triggers, it is found
that a little wider for away-side correlations trigged by any
hadrons than the one trigged by $\phi$ or $\Omega$, while it keeps
almost same width for near-side peaks. The former may indicate
that  $s$ or $\overline{s}$ quarks has weak partonic interactions
with $u$ or $d$ quarks, which results in a narrower width of
away-side peaks for $\phi$/$\Omega$-trigged correlation. Of
course, we selected a little wider $p_T$ range for $\phi$ and
$\Omega$ triggers due to the limited statistics, which may have a
small effect on the differences of di-hadron correlations. For the
peripheral collisions, right panel of Fig.~\ref{dihadron}
demonstrated that the Mach-like cone peaks vanish and only mono
peak exists in away-side which is also consistent with the
experimental observation. The matter in peripheral collisions is
more dilute so that the associated particles can be easily
traverse the system directly. In this case, we have a single peak
in the backward of the trigged particles. Based on the same reason
for the central collisions, the associated particles trigged by
$\phi$ and $\Omega$ have a narrower width for the away-side peak.
From the above simulations, some differences in di-hadron
correlation functions between $\phi$/$\Omega$ trigger and the
normal particles trigger are predicted, which deserves to be
explored in  the data.

\vspace{-0.2cm}
\section{CONCLUSIONS}
\vspace{-0.2cm}

STAR and PHENIX have measured wealth of data which help ones to
learn the properties of $\phi$-meson production and the behavior
of elliptic flow $v_{2}$ at RHIC. The measurement of $\phi$-meson
transverse momentum spectra as a function of centrality for Au+Au
collisions at $\sqrt{s_{NN}}$ = 200 GeV shows that for peripheral
collisions, the spectra has a deviation of exponential shape
because of the distinct power-law tail and are best described by a
Levy function while for the central data, the power-law tail is
suppressed and it can be well described by both Levy or
exponential function. In other words, clear central suppression of
transverse momentum spectra has been observed. Collisions among
different systems (Au+Au, d+Au, p+p at $\sqrt{s_{NN}}$ = 200 GeV)
indicate that the central suppression only exists in the Au + Au
systems. The SATR measurement of nuclear modification factor
($R_{CP}$) conclusively demonstrated that the $\phi$-meson
$R_{CP}$ is much similar to the $R_{CP}$ of $K_{S}^{0}$ than
$\Lambda$, which confirms the grouping of hadron $R_{CP}$
observable at RHIC is a baryon-meson type dependence but not a
mass type dependence. The first measurement of $\phi-meson$
elliptic flow $v_{2}$ from STAR shows that $\phi$ flows as strong
as other particle. At low $p_{T}$ ($<2$ GeV/c) region, $v_{2}$ of
$\phi$ is consistent with hydrodynamical behavior of strange
quarks in the early partonic stage while at intermediate region (2
GeV/c $<p_{T}<5$ GeV/c), it favors the number of constituent quark
scaling. The sizeable $v_{2}$ from the data is a strong signal of
partonic collectivity at RHIC. In addition, a Mach-like cone
picture in central collisions is demonstrated for $\phi$ or
$\Omega$ trigged di-hadron correlation with AMPT model and it
reveals a narrower width of away-side peaks than the normal
particles trigged di-hadron correlation function, which deserve to
be explored experimentally in near future.

\vspace{0.2cm}

\hspace{-1.0cm} Acknowledgments: We would like to thank J. H.
Chen, G. L. Ma, X. Cai and S. Zhang from SINAP, H. Z. Huang, J. G.
Ma from UCLA, N. Xu and S-L. Blyth from LBNL for useful
discussions. The work was supported in part by the Shanghai
Development Foundation for Science and Technology under Grant
Numbers 05XD14021, the National Natural Science Foundation of
China under Grant No 10328259 and 10535010.

\end{document}